# Mid-infrared feed-forward dual-comb spectroscopy at 3 µm

Zaijun Chen [1], Theodor W. Hänsch [1,2], Nathalie Picqué [1]

1. Max-Planck-Institut für Quantenoptik, Hans-Kopfermann-Straße 1, 85748 Garching, Germany
2. Ludwig-Maximilians-Universität München, Fakultät für Physik, Schellingstr. 4/III, 80799 München, Germany

**Abstract**
Mid-infrared high-resolution spectroscopy has proven an invaluable tool for the study of the structure and dynamics of molecules in the gas phase. The advent of frequency combs advances the frontiers of precise molecular spectroscopy. Here we demonstrate, in the important 3-µm spectral region of the fundamental CH stretch in molecules, dual-comb spectroscopy with experimental coherence times between the combs that exceed half-an-hour. Mid-infrared Fourier transform spectroscopy using two frequency combs with self-calibration of the frequency scale, negligible contribution of the instrumental line-shape to the spectral profiles, high signal-to-noise ratio and broad spectral bandwidth opens up novel opportunities for precision spectroscopy of small molecules. Highly multiplexed metrology of line shapes may be envisioned.

Mid-infrared frequency combs with unprecedented mutual coherence times open the important spectral region around 3 µm to multiplexed gas phase molecular spectroscopy with high precision and negligible instrumental line width. The spectrum of a laser frequency comb [1] spans a broad bandwidth and is made up of phase-coherent evenly spaced narrow lines, the frequency of which may be known within the accuracy of an atomic clock. Frequency combs find numerous new applications. In spectroscopy over broad spectral bandwidths, they are used to directly interrogate the transitions of atomic and molecular species. One of the techniques of broadband frequency comb spectroscopy, dual-comb spectroscopy [2], currently attracts significant interest. Dual-comb spectroscopy is a technique of linear [3-6] and nonlinear [7, 8] Fourier transform spectroscopy without moving parts. In most implementations of linear spectroscopy, a frequency comb source interrogates an absorbing sample and is heterodyned against a second comb of slightly different line spacing. The measured time–domain interference leads to a spectrum through the harmonic analysis provided by the Fourier transformation.

The development of dual-comb spectroscopy in the regions of interest to molecular spectroscopy, such as the mid-infrared range where most molecules have strong vibrational transitions, turns out to be rather challenging for several reasons.

Novel laser technology is still emerging and many different approaches to mid-infrared frequency comb generation [9] have been and are being explored. They include quantum cascade [10] and interband cascade [11] lasers, microresonators [12, 13], novel solid-state [14] or doped-fiber [15] gain media, improved approaches to nonlinear frequency conversion [16] benefiting optical parametric oscillation, difference frequency generation and spectral broadening in nonlinear waveguides [17, 18].

Furthermore, the technical requirements for a dual-comb spectrometer are more challenging than those for a metrological frequency comb. A dual-comb interferometer requires that the optical delay between the pairs of interfering pulses is controlled, from pulse pair to pulse pair, within interferometric accuracy. Timing fluctuations in the attosecond range can be detrimental. In the near-infrared domain, coherence between the two combs at most reached the order of 1 second in different experimental realizations [19-21]. The measurement time can be extended to several minutes or hours by more or less complex techniques of phase correction, that are implemented through analog [22], digital [23] or *a posteriori* computer [24] treatment. Maintaining coherence in the mid-infrared domain is even more difficult than in the near-infrared region, because e.g. of the more complex laser systems and of the difficulty to design continuous-wave mid-infrared lasers with a narrow line-width, that could serve as optical references for the combs.

Finally experimenting in the mid-infrared region is generally more demanding than in the visible or near-infrared region because the optics and photonics technology is not as advanced.

Consequently many reports of mid-infrared dual-comb spectroscopy, although pointing to an intriguing potential, have remained at a stage of promising proof-of-principle demonstrations. The various laser systems developed lately have led to a variety of implementations [14, 25-30]. Moreover, the first demonstrations with frequency combs of large line spacing, such as microresonators [31] and quantum cascade lasers [32], highlight intriguing opportunities in time-resolved spectroscopy for physical chemistry in the condensed phase. Nevertheless, quantum cascade [33] and interband [34] cascade lasers, difference-frequency generation using erbium doped fiber lasers [3] and electro-optic modulators [6] have produced, over narrow spectral bandwidths, high-quality spectra with resolved comb lines. Recently the control of degenerate optical parametric oscillators has also been remarkably advanced [35], while a combination of difference-frequency combs stabilized in the near-infrared and of real-time digital correction proved successful in producing [4] spectra with resolved comb lines over a broad spectral bandwidth.

Not long ago, we have devised [5] a new scheme of dual-comb interferometry that reports unparalleled coherence times, more than three orders of magnitude longer than the state-of-the-art. The initial demonstration has been performed in the near-infrared region with erbium-doped fiber oscillators spectrally broadened in nonlinear fibers. The near-infrared interferometer, based on feed-forward stabilization of the relative carrier-envelope offset frequency, does not require any types of phase correction for continuous averaging of interferograms up to half-an-hour, resulting in greater experimental simplicity and reduction of possible artifacts. Here we explore the extension of our scheme to broadband highly accurate dual-comb spectroscopy in the mid-infrared 3-µm range, where the fundamental CH, NH, OH stretches in molecules are found.

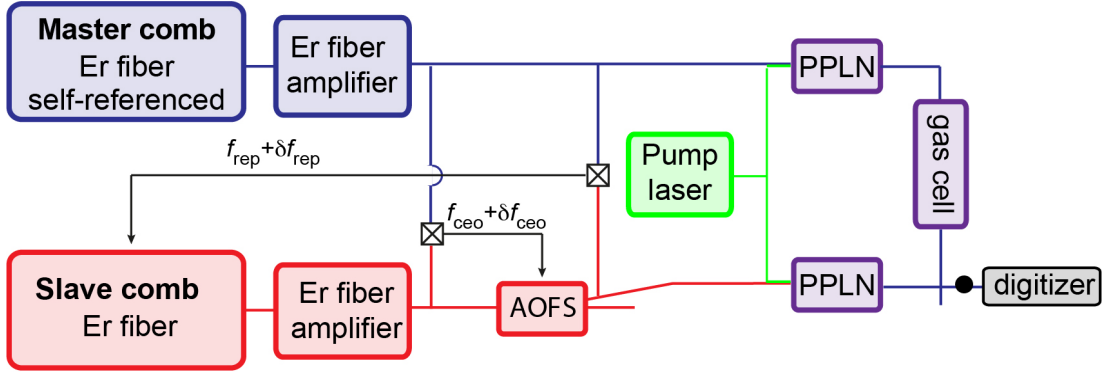

**Figure 1.** Experimental set-up for mid-infrared broadband dual-comb spectroscopy using feed-forward control of the relative carrier-envelope offset frequency. AOFS: acousto-optic frequency-shifter, PPLN: periodically-poled lithium niobate crystal.

Our 3-μm dual-comb laser system (Fig.1) is based on difference frequency generation of near-infrared femtosecond erbium-doped fiber lasers.
The near-infrared dual-comb system has already been described in [5]. We summarize here its main features for the clarity of the description. Two amplified mode-locked erbium-doped fiber laser systems emit trains of pulses at a repetition frequency of about 100 MHz. Their spectrum is centered at an optical frequency of about 190 THz and it spans more than 20 THz. The repetition frequency $f_{rep}$ and the carrier-envelope offset frequency $f_{ceo}$ of the first fiber laser system (called master comb) are stabilized against the radio-frequency clock signal of a hydrogen maser, and all electronic instruments in our set-up are synchronized to this clock signal. In the experiments described below, we choose $f_{rep}$= 100.00 MHz and $f_{ceo}$=20.0 MHz and both radio-frequencies are counted during the recording of the interferograms. The second fiber laser system, called slave comb, has a slightly different repetition frequency $f_{rep}+\delta f_{rep}$ (with $\delta f_{rep} \ll f_{rep}$) and a different carrier-envelope offset frequency $f_{ceo}+\delta f_{ceo}$. Here, we set $|\delta f_{rep}|$=130 Hz, which leads to an optical free spectral range of 38 THz. For interferometry measurements, the optical retardation between pairs of interfering pulses must be accurately controlled during the duration of the measurement. Our technique is based on feed-forward control of the relative carrier-envelope offset frequency: an acousto-optic frequency shifter, at the output of the slave laser system, keeps $\delta f_{ceo}$ constant over arbitrary times by frequency translating, with a fast response time (about 550 ns), all the comb lines of the slave comb. A slow feedback loop (<1 kHz bandwidth) adjusts the length of the cavity of the slave comb in order to maintain a fixed difference in repetition frequencies $\delta f_{rep}$. The signals that enable to monitor the relative fluctuations between the combs are two beat notes between two pairs of optical comb lines, one from each comb. Two continuous-wave lasers, emitting at 189 THz and 195 THz, respectively, serve as intermediate oscillators to produce the two beat notes. For the spectroscopy set-up, the output of the master comb and that of the first-order diffracted beam of the acousto-optic frequency shifter of the slave comb have an available average power of about 250 mW each.
Each near-infrared signal comb is converted to a mid-infrared idler comb by difference frequency generation in a periodically poled lithium-niobate (PPLN) crystal of a length of 3 mm and with 7 poling periods of about 30 µm. The pump beam is provided by a continuous-wave laser that emits at $f_{\text{pump}} \approx 281.8$ THz (1063.8 nm) and that is phase-locked to a line of the master comb. Its free-running linewidth at 100 µs is specified to be 50 kHz. For frequency calibration of the spectra, we count the absolute frequency of the pump laser $f_{\text{pump}}$, that of the 189-THz continuous-wave intermediate-oscillator laser and the radio-frequency parameters of the two combs during the measurement of the interferograms. The pump is split into two beams. For each, an average power of 2W is combined with 250 mW of signal beam by means of a dichroic mirror and each sumperimposed beam is focused onto a mixing PPLN crystal. We generate an idler comb spanning up to 8.2 THz - a span limited by the phase matching bandwidth of the PPLN crystal– with an average power up to 90 µW. Such a technique of difference frequency comb generation has already been harnessed in dual-comb spectroscopy with erbium fiber lasers [3] and with electro-optic modulators [6]. Mixing a continuous-wave laser and an ultra-short pulse is a fairly inefficient process; the simplicity of its implementation and the fact that, due to detector nonlinearities, we do not require high mid-infrared powers, justify though that we retained this scheme. The efficiency could be much increased with a build-up cavity for the pump beam and by introducing a pulse-stretching chirp for the pulses of the idler combs. The repetition frequency of the idler comb is the same as that of the signal comb. The carrier-envelope offset frequency of the idler comb is shifted by an amount $f_{\text{pump}}$ [modulo $f_{\text{rep}}$]. As the same pump laser is used for the generation of the two idler combs, the two carrier-envelope offset frequencies are shifted by the same amount and their difference remains $\delta f_{\text{ceo}}$. The center frequency of the idler comb is adjustable between 82 THz and 100 THz by changing the temperature and the poling period of the PPLN crystal. After each PPLN crystal, an optical long-wavelength-pass filter filters out the residual pump and signal beams. The idler master comb beam passes through a single-pass absorption cell of a length of 70 cm. The master and slave idler beams are combined on a pellicle beam-splitter. One of the outputs of the beam-splitter is focused onto a fast thermoelectrically cooled HgCdTe photodetector. To avoid detector nonlinearities, that generate distortions and systematic effects in the spectra, the total average power at the detector is kept weaker than 40 µW. The electric signal is amplified and filtered. It is sampled synchronously to the master comb repetition frequency, $f_{\text{rep}}$ = 100 MHz, by a fast digitizer. To keep the individual interferograms in phase and enable their averaging in the time domain, we set $\delta f_{\text{ceo}} = 0$ [modulo $\delta f_{\text{rep}}$]. The time domain interference signal is Fourier transformed to reveal the spectrum. Integration times of several minutes are required to get good signal-to-noise ratios over a broad spectral bandwidth. We average the raw interferograms in the time domain by simply adding them up. We do not perform any numerical corrections (like phase correction) to the interferograms or to the spectra. The displayed transmission spectra are the amplitude of the complex Fourier transform of the averaged interferograms while the dispersion spectra are their phase.

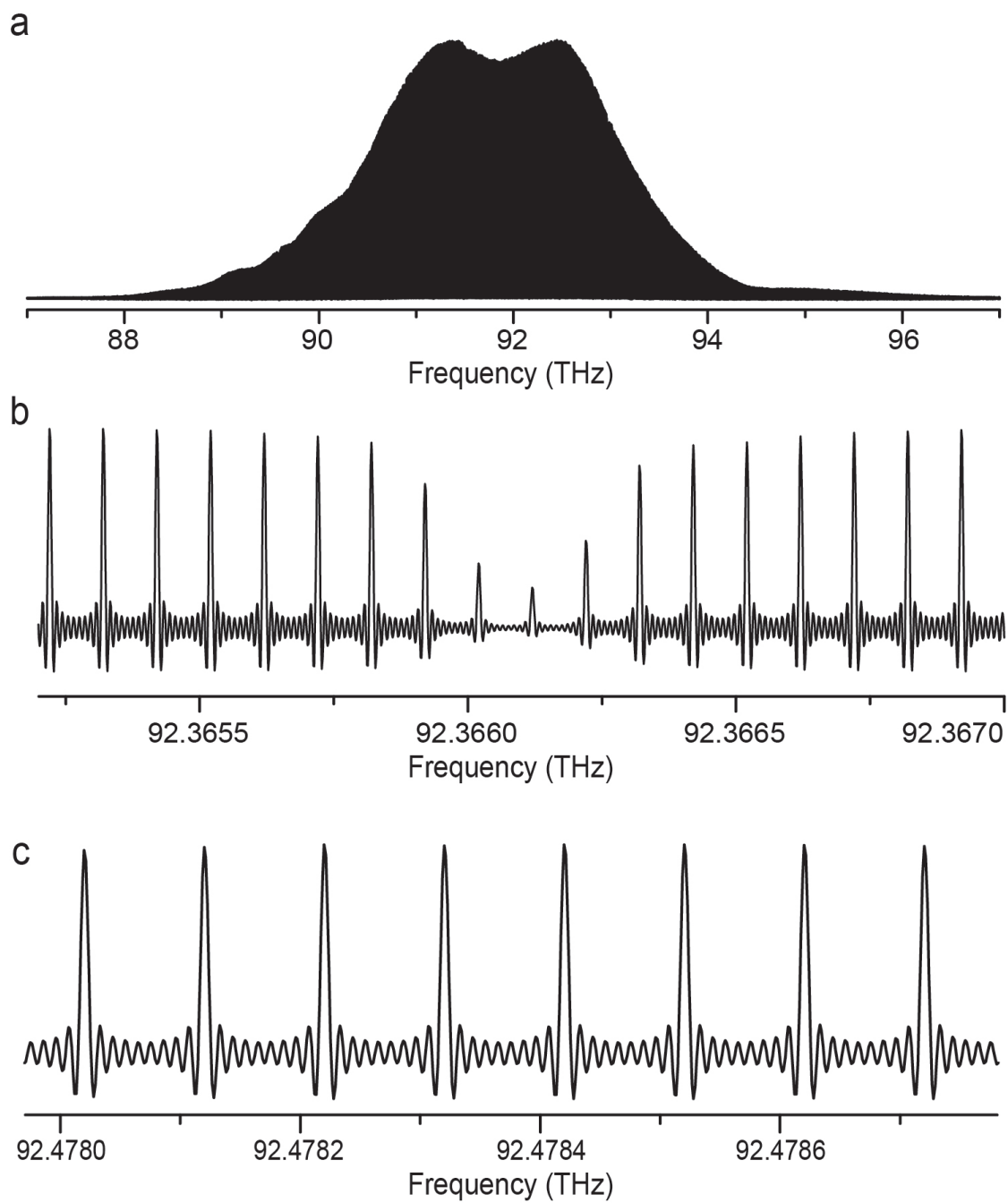

**Figure 2.** Experimental dual-comb spectrum with resolved comb lines., recorded within 1742 seconds. **a.** Entire apodized spectrum: 82000 individual comb lines spanning 8.2 THz are resolved. **b.** Detail of the unapodized representation of a. showing the ($J'$=13,$K_a'$=1,$K_c'$=13)-($J''$=14,$K_a''$=0, $K_c''$=14) line in a *P*-branch of the $\nu_9$ band of $^{12}C_2H_4$ sampled by the comb lines of 100-MHz spacing. **c.** Magnified unapodized representation of a. showing eight individual comb lines with their cardinal sine instrumental line shape.

Figure 2a displays a transmission spectrum with 82000 resolved comb lines, resulting of an interferogram averaged during 29 minutes (1742 s). For illustrating the resolved comb lines, interferometric sequences of 0.146 s including 19 individual interferograms are time averaged and the resulting interferogram is Fourier transformed. In Fig. 2a, the spectrum is shown over the entire useful spectral span of 8.2 THz, from 88 THz to 96.2 THz, with triangular apodization of the interferogram. Fig. 2b shows an expanded view of the unapodized spectrum, illustrating that the ro-vibrational transitions of the absorbing sample, here ethylene $C_2H_4$ in natural abundance, with a Doppler full-width at half-maximum of 215 MHz at 296 K are satisfactorily sampled by the comb lines of 100-MHz spacing. The ethylene pressure is 147 Pa and the absorption path length is 70 cm. The individual comb lines (Fig. 2c) have the expected line shape, a cardinal sine due to the finite measurement time, which convolves the comb beat notes of negligible width. The full-width at half-maximum of the observed comb lines is 6.8 Hz in the radio-frequency domain, which corresponds to the Fourier transform limit. In Fig. 2c, the frequency scale is converted to the optical scale, thus the width of the comb lines appears as 5.3 MHz. The measurement time is similar to that in our recent near-infrared spectra [5] that use the same feed-forward technique and we observe no degradations in the retrieved instrumental line shapes due to the operation in the mid-infrared region.

The spectrum sampled at exactly the comb line positions reveals (Fig.3a) the absorption by ethylene. Ethylene $C_2H_4$ is a near-prolate planar asymmetric top molecule. Our span covers the region of the $v_9,v_{11}$-stretching dyad of $^{12}C_2H_4$, which includes the $v_2+v_{12}$, $2v_{10}+v_{12}$ and $v_9+v_{10}$ cold bands. Figures 3b and 3c illustrate details of the transmission and dispersion spectra. Figure 3b provides an expanded view of a part of the $v_9$ band, which corresponds to a CH2 stretching mode. The strong lines are unresolved doublets in the P branch of the ($K'_a=5$, $\Delta K_a = K'_a - K''_a = -1$, $\Delta K_c = K'_c - K''_c = +/-1$) series. $K_a$ (respectively $K_c$) is the quantum number for the projection of the rotational angular momentum (excluding electron and nuclear spin) onto the inertial axis of smallest moment (respectively largest moment). Prime and double primes refer, respectively, to upper and lower states, respectively. The ssignments are taken from the report [36]. The weaker Q-branch of the ($K'_a=7$, $\Delta K_a=-1$, $\Delta K_c = K'_c - K''_c = 1$) series around 91.27 THz is well resolved. Figure 3c expands another area, where the intensity of the spectral envelope of our dual-comb system is significantly weaker, showing the region of Q-branch of the ($K'_a=7$, $K''_a=6$, $\Delta K_c=-1$) lines in the $v_9$ band. The signal-to-noise ratio at 92.3 THz peaks at 1275 for a measurement time of 1742 s, corresponding to 30.5 $s^{-1/2}$. The average signal-to-noise ratio is 570 (or 13.6 $s^{-1/2}$) across the full span of 8.2 THz, and the corresponding figure of merit, which is given by the product of the average signal-to-noise ratio per unit square root of measurement time and the number of spectral elements 82000, is $1.1 \times 10^6$ $s^{-1/2}$. Such a figure of merit is comparable to those reported in the same spectral region for other experiments of narrowband [3, 6] or digitally-corrected a broadband [4] dual-comb spectroscopy. As shown in Figure 4, when the interferograms are averaged in the time domain, the average signal-to-noise increases as the square root of the measurement time without any indication that the behavior reaches saturation. This demonstrates that the experimentally achieved interferometric coherence reaches half an hour. Such duration, obtained by time-domain averaging of the interferograms without any other treatment, is significantly longer than what is achieved with the state-of-the-art dual-comb systems, where an experimental mutual coherence time of 1 second used to be an excellent figure. Furthermore, it provides convincing evidence of the power of our technique of feed-forward dual-comb interferometry in the molecular fingerprint region and it renders superfluous any types of phase corrections or other numerical manipulations of the experimental data.

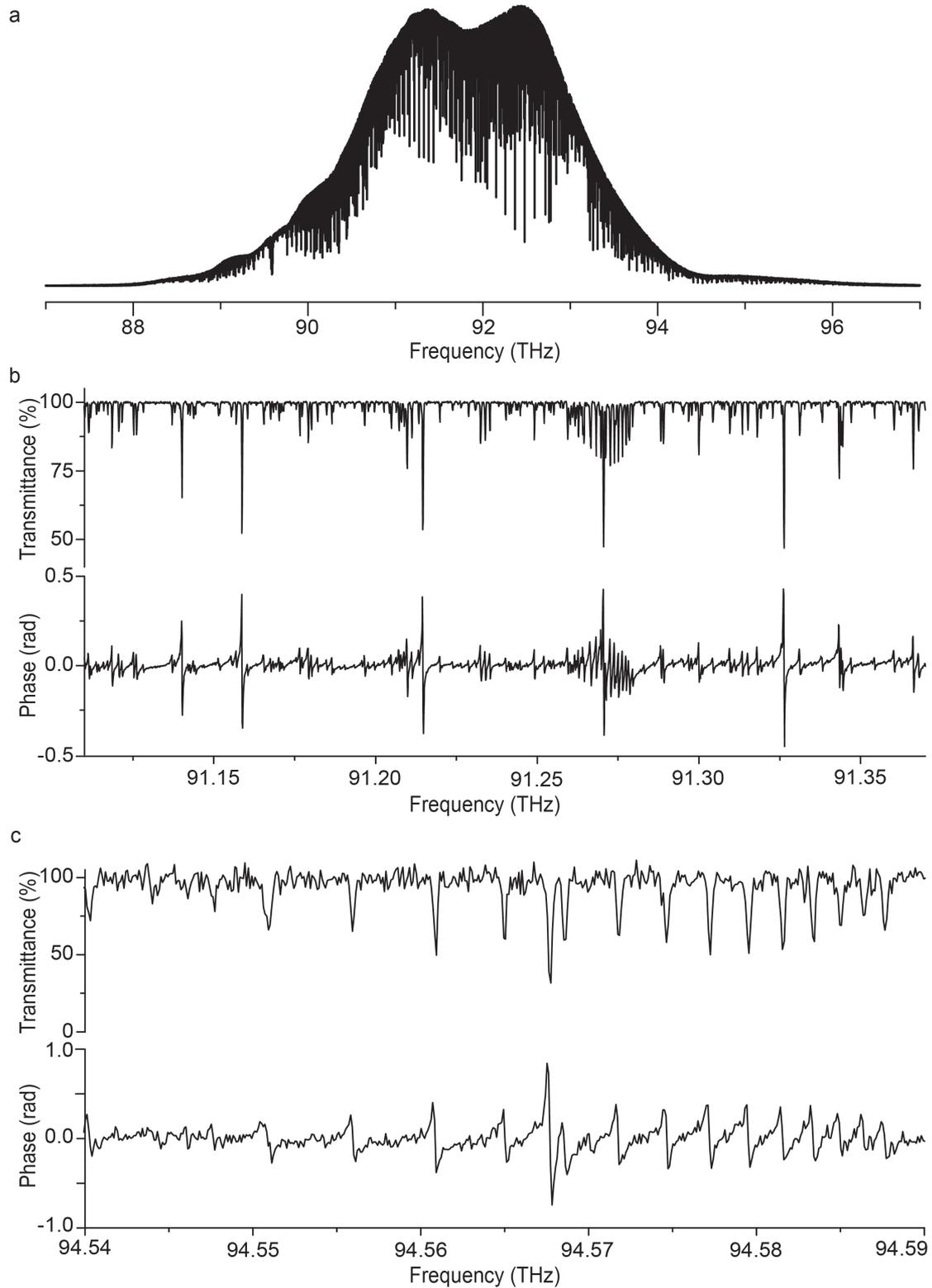

**Figure 3.** Experimental dual-comb spectrum of ethylene sampled at exactly the comb line spacing of 100 MHz in the region of 92 THz. **a.** The amplitude of the Fourier transform provides the transmission spectrum **b.** Magnified representation of the transmittance (top) and dispersion (bottom) spectra around 91.25 THz. Note that the transmittance y-axis does not go to zero. **c.** Magnified representation of the transmittance (top) and dispersion (bottom) spectra around 94.56 THz.

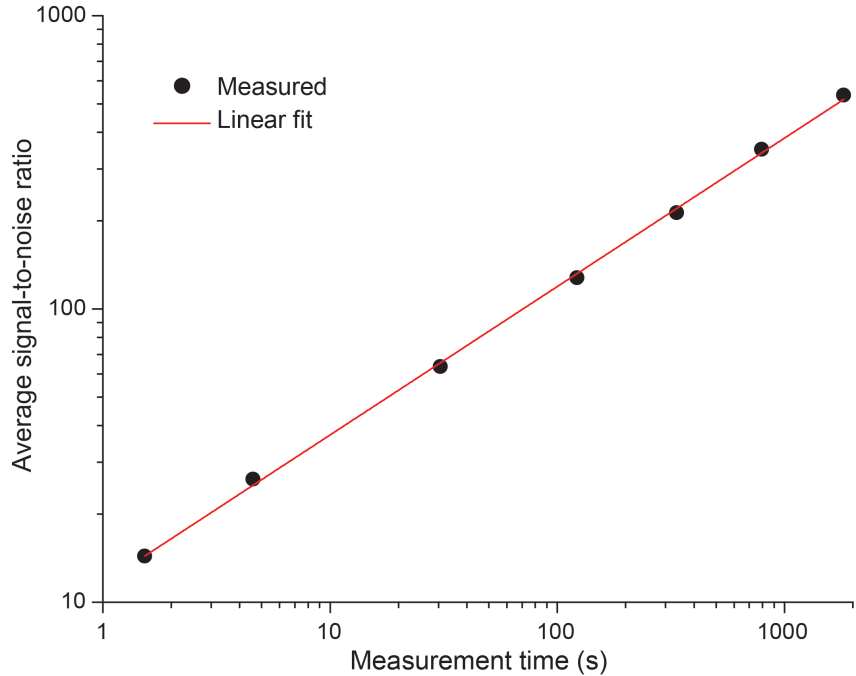

**Figure 4.** Evolution of the average signal-to-noise ratio with the measurement time in an ethylene spectrum. The fit of a line to the experimental data shows a slope of 0.506 (5) indicating that the signal-to-noise ratio shows dependence with the square root of the measurement time. The spectrum corresponding to a measurement time of 1742 s is shown in Fig. 2 and Fig. 3.

Adjusting the temperature and poling period of the PPLN crystals provides a slight tunability of the idler spectrum and makes it possible to optimize the spectrum to the transitions of different molecules, as illustrated in Fig. 5 which displays the region of the *P*-branch of the $\nu_3$ band of acetylene. The strongest lines in Figure 5b are assigned to the $\nu_3$ and $\nu_2+(\nu_4+\nu_5)_+^0$ cold bands of $^{12}C_2H_2$, while the weaker ones belong to the $\nu_3+\nu_4^1-\nu_4^1, \nu_3+\nu_5^1-\nu_5^1$, $\nu_2+(2\nu_4+\nu_5)^1 \text{II} -\nu_4^1$, $\nu_2+(\nu_4+2\nu_5)^1\text{II}-\nu_5^1$ hot bands (their rotational assignment may be found in [37]) and to the $\nu_3$ band of $^{13}C_2H_2$. The notation for the bands follows that proposed by [38] and widely adopted [37]. In particular, the number in superscript of the vibrational levels $\nu_4$ and $\nu_5$ (which correspond to the cis- and trans-degenerate bending modes, respectively) is the quantum number $\ell = |\ell_4+\ell_5|$, where $\ell_i$ is the quantum number of vibrational angular momentum associated with the degenerate bending mode i. The + in subscript is the symmetry type of the involved $\Sigma$ state. The roman number II provides the rank of the level, by order of decreasing energy, inside the ensemble of states of identical vibrational symmetry coupled by $\ell$-type resonances. The pressure of acetylene in natural abundance is 10.7 Pa and the absorption path length is 70 cm. The spectrum results from an interferogram averaged during 2050.2 s (34.2 min) in the time domain. The signal-to-noise ratio at around 98.4 THz is 1360, corresponding to 30 s$^{-1/2}$. The average signal-to-noise ratio, across the span of 8 THz, is about 500 (or 11 s$^{-1/2}$). The signal-to-noise ratio increases as the square root of the measurement time.

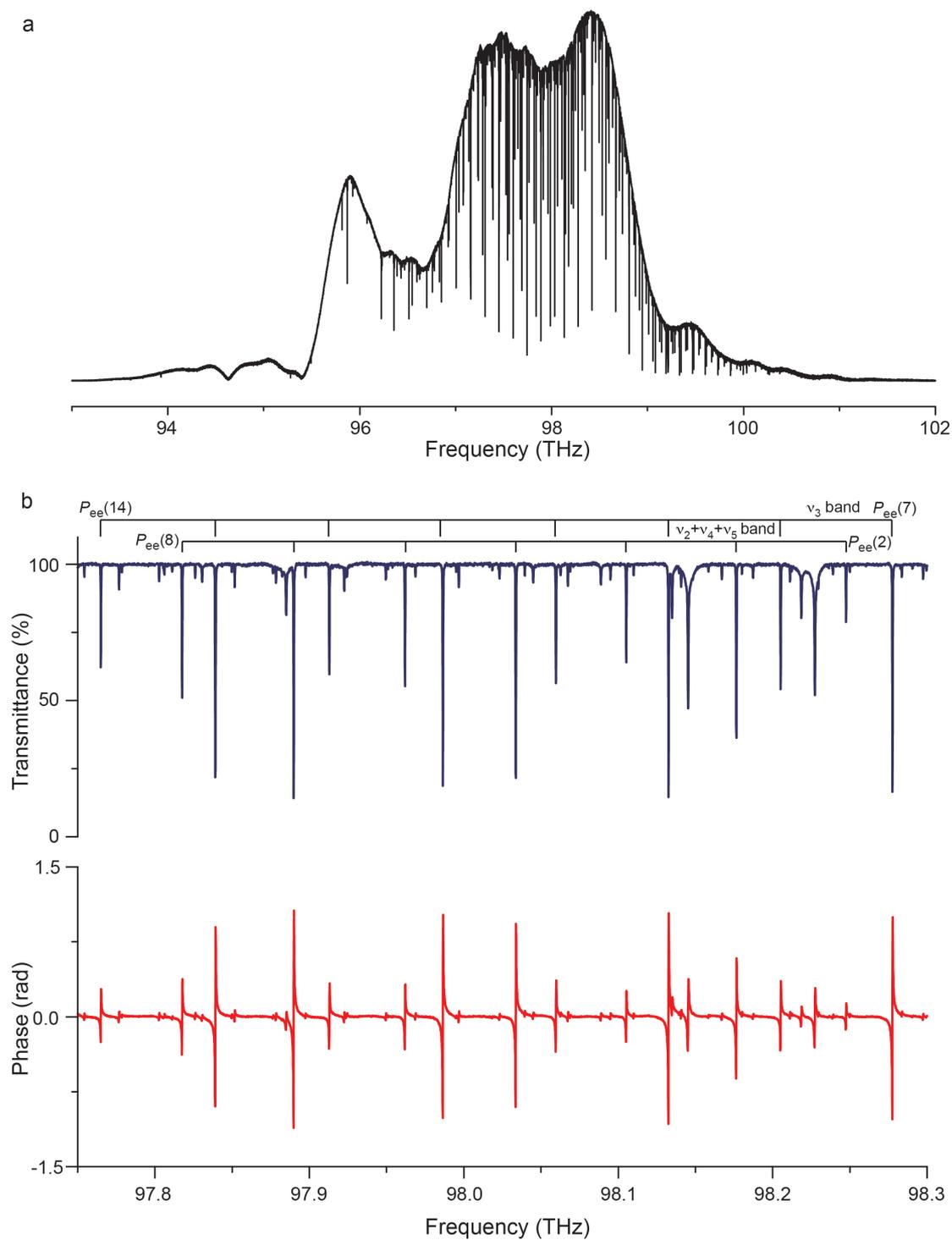

**Figure 5.** Experimental spectrum of acetylene in the region of the $\nu_3$ band with a resolution of 100 MHz. **a.** The entire spectrum is measured within 2050.2 s. **b.** Transmittance and dispersion spectra in the region of the *P*-branch of the $\nu_3$ band of $^{12}C_2H_2$. The assignments for the strongest lines are provided in the figure. The lines with a broader pedestal are due to atmospheric water vapor, outside the cell, in the beam path.

As a preliminary analysis of our molecular spectra we have fitted the profiles and determined the positions of a limited number of lines amongst the hundreds present in each crowded spectrum. Figure 6 exemplifies the results of the adjustments of Doppler profiles to three experimental lines in the spectrum of ethylene shown in Fig. 3. The standard deviation of the "observed-fitted" residuals is 0.12% and no systematic residual signatures are distinguishable at the noise level. We compare the positions of about 250 selected isolated lines of $^{12}C_2H_4$ to those available in the literature. The retrieved positions in our line selection present an average statistical uncertainty of 0.8 MHz with a standard deviation of 0.4 MHz. An extensive linelist is available in the 2016 HITRAN database [39], which reproduces the report [36]. In this report [36], the precision of the line-position measurements is given to +/- 20 MHz and the accuracy is said to be "slightly worse". The average value of the difference between the positions in [36] and ours is 0.3 MHz and the standard deviation is 1 MHz. About sixty of our selected lines, in the $v_{11}$ band, were also measured in [40] with an accuracy of +/- 18 MHz. The average value of the difference between the positions in [40] and ours is 1.0 MHz and the standard deviation is 3.2 MHz. This indicates that precision position measurements of Doppler-broadened transitions may be achieved with our set-up. The accuracy of the frequency scale is granted by the direct calibration to the hydrogen maser. Special care was also taken to operate the photo-detector within its linearity range in order to avoid systematic shifts due to detector nonlinearities. However, biases at the sample, such as pressure shifts, have not been evaluated yet and will be the object of future investigations.

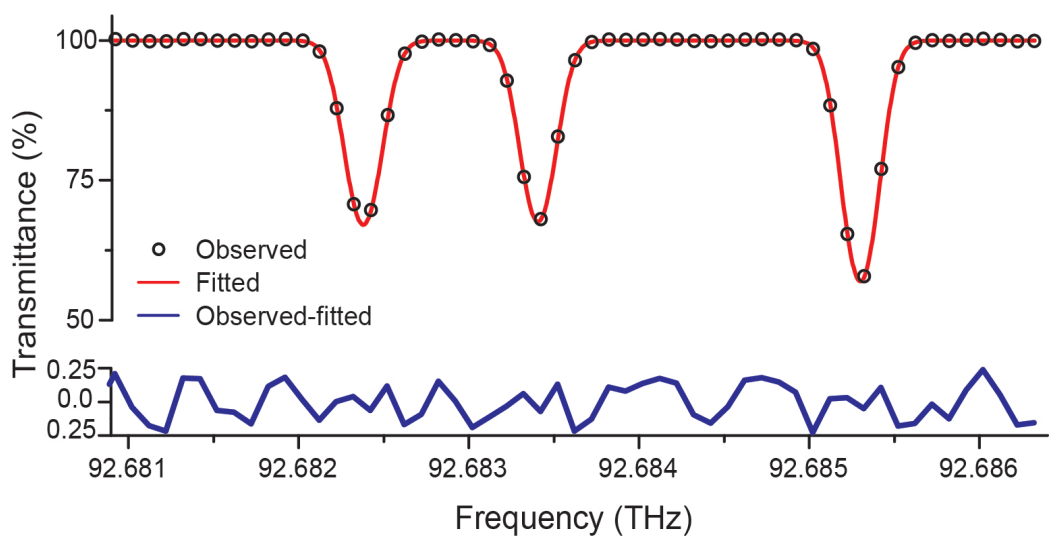

**Figure 6.** Portion of the transmittance spectrum of $^{12}C_2H_4$ (black open dots, labeled observed), magnifying three rovibrational lines. The transmittance y-scale in the spectrum stops at 50%. The experimental profile is satisfactorily fitted by a Doppler line shape (red line, labeled fitted). The standard deviation of the residuals observed – fitted is at the noise level. Note that the y-scale of the residuals is magnified compared to the y-scale of the part showing the spectrum.

Here we demonstrate the first self-referenced broadband mid-infrared dual-comb spectroscopy, with resolved comb lines and negligible contribution of the instrumental lineshape, that does not require any types of phase correction. We experimentally achieve mutual coherence times that far exceed the state-of-the-art. Meanwhile the instrumental set-up for coherence control is relatively simple. The implementation of real-time phase correction algorithms, that reconstruct the coherence, requires complex computer hardware and programming. Compared to these, our technique is straightforward. Compared to locking the combs to optical references, our technique does not require reference cavities (nor other means of achieving a continuous-wave laser with a line-width of a hertz or below). As already discussed in [5], phase correction is not always implementable and it may otherwise induce subtle artifacts detrimental to precision measurements. Thus an instrument that does not require any data corrections will be useful to broadband precision spectroscopy and it will help ascertaining the precision frontiers and identifying tiny systematic instrumental effects.

Our spectrometer provides self-calibration of the frequency scale directly within the accuracy of an atomic clock, as well as a negligible contribution of the instrumental line shape to the experimental profile. As the achieved mutual coherence is excellent, the optical comb lines of the radio-frequency-referenced master comb that interrogate the sample have the strongest contribution to the instrumental profiles. The width of the optical comb lines is about 100 kHz at an integration time of 1 minute, thus three orders of magnitude narrower than the Doppler width. Therefore, the intrinsic molecular line shape can directly fit the line profiles. Even for spectroscopy of Doppler-broadened profiles at room temperature, improved line parameters -not limited to positions and shifts- may be determined. To date, most of the precise spectroscopic individual line parameters retrieved from high-resolution infrared spectra have derived from Michelson-based FT spectroscopy with an incoherent light source, where the wavenumber-scale calibration relies on the presence of molecular lines, accurately measured by other means, that can serve as wavenumber standards in the spectra. The instrumental line-shape in Michelson-based FTS has a width generally of the same order of magnitude as the intrinsic width of the observed lines and has to be included in the fits. The use of an incoherent light source in Michelson-based FT spectroscopy enables recordings over extended spectral bandwidths that are still out of reach with frequency comb spectroscopy. Over limited bandwidths, which are commonly used in Michelson-based FTS for signal-to-noise ratio improvement, coherent light sources of high brightness represent [41, 42] a significant increase in signal to noise ratio (or decrease in measurement time). Our system benefits from this high brightness and adds the advantages of the accuracy of the frequency scale, of the precision of the line profile determination and of the absence of moving parts.

One of the limitations of the dual-comb system presented here is the low average power (90 µW) of each mid-infrared comb. For spectroscopy in a single-pass cell or in a multipass cell with a small number of reflections, the achieved average power is appropriate, as the total power on the detector needs to be kept lower than 30-40 µW to avoid detector nonlinearities. For laboratory spectroscopy over long paths in a multipass cell or in a high finesse cavity [43], the power would likely be too low. In our configuration though, the pump photons from the continuous wave laser are entirely depleted. By chirping the signal pulses to several picoseconds, we increase the conversion efficiency and the average power by one order of magnitude, which would be sufficient for multipass cells. For experiments requiring significantly higher power, other comb sources, such as optical parametric oscillators [35] or systems based on difference frequency generation between ultra-short pulses [4, 26, 44, 45], would have to be harnessed. The feed-forward relative control could be implemented on the mid-infrared beams. This still has to be experimentally demonstrated, but is technically achievable. With our set-up, larger tunability of the central frequency could be straightforwardly achieved by spectrally broadening the output of the erbium oscillators in nonlinear waveguides and/or by using a tunable continuous-wave laser as the pump in the difference frequency generation.

The 3-μm region is rich in strong molecular transitions, belonging e.g. to fundamental bands of many small and large organic, nitrogen-containing and oxygen-containing molecules, of relevance to fundamental and applied spectroscopy. Future work will exploit our feed-forward dual-comb spectrometer for precise spectroscopy of molecular line shapes.

**Acknowledgments**

Support by the Munich Center for Advanced Photonics and by the Carl-Friedrich-von-Siemens Foundation is gratefully acknowledged.